\documentclass[conference]{IEEEtran}

\ifCLASSINFOpdf
\else
\fi

\usepackage{amsmath}
\usepackage{graphicx}
\usepackage{caption}
\usepackage{tabularx,multirow,booktabs,lipsum}
\usepackage{subfig}
\usepackage{color}
\usepackage{float} 
\usepackage{paralist} 
\usepackage[T1]{fontenc}
\usepackage{currvita}
\usepackage{epstopdf}
\usepackage{amsmath}
\usepackage{amsfonts}
\usepackage{amssymb}

\begin{document}

\title{Anti-Jamming Strategy for Distributed Microgrid Control based on Power Talk Communication}

\author{Pietro Danzi, Marko Angjelichinoski, \v{C}edomir Stefanovi\'c, Petar Popovski\\
Department of Electronic Systems, Aalborg University, Denmark,\\ 
Email:\{pid,maa,cs,petarp\}@es.aau.dk }

\maketitle

\begin{abstract}
In standard implementations of distributed secondary control for DC MicroGrids (MGs), the exchange of local measurements among neighboring control agents is enabled via off-the-shelf wireless solutions, such as IEEE 802.11.
However, Denial of Service (DoS) attacks on the wireless interface through jamming prevents the secondary control system from performing its main tasks, which might compromise the stability of the MG.  
In this paper, we propose novel, robust and secure secondary control reconfiguration strategy, tailored to counteract DoS attacks.
Specifically, 
upon detecting the impairment of the wireless interface, the jammed secondary control agent notifies its peers via a secure, low-rate powerline channel based on Power Talk communication.
This triggers reconfiguration of the wireless communication graph through primary control mode switching, where the jammed agents leave the secondary control by switching to current source mode, and are replaced by non-jammed current sources that switch to voltage source mode and join the secondary control.
The strategy fits within the software-defined networking framework, where the network control is split from the data plane using reliable and secure side power talk communication channel, created via software modification of the MG primary control loops.
The simulation results illustrate the feasibility of the solution and prove that the MG resilience and performance can be indeed improved via software-defined networking approaches.
\end{abstract}


\IEEEpeerreviewmaketitle

\section{Introduction}
\label{intro}

The introduction of affordable and efficient Distributed Energy Resources (DERs) based on renewable generation stimulated the development of small, localized and self-sustainable direct current (DC) MicroGrids (MG).
MGs are regulated and optimized through hierarchical control system, comprising \emph{primary}, \emph{secondary} and \emph{tertiary} control loops, implemented in the power electronic converters that interface DERs to the DC distribution infrastructure \cite{ref1}.
The primary control maintains the electric variables within predefined margins to guarantee MG stability; it is implemented in a decentralized manner, relying only on local measurements as feedback.
The upper (secondary/tertiary) controls restore the voltage to its nominal reference and optimize the power flow; they are commonly implemented in either centralized or distributed manner, via consensus algorithm and require exchange of local information among neighboring control agents \cite{ref13}.

In standard implementations, low cost, off-the-shelf wireless communication solutions, such as IEEE 802.11 WiFi standards, are used to support the upper level control \cite{ref3}. 
It is well known that the WiFi systems are vulnerable to specific types of security attacks by malicious entities; Denial-of-Service (DoS) is an example of such attack, which can be easily conducted through jamming \cite{ref9}.
The effects of DoS attacks on distributed secondary control have been investigated in \cite{ref4}.  
It was shown that DoS attacks lead to prolonged communication outages, splitting the communication graph and preventing the secondary controllers from restoring the voltage and keeping the current sharing balance.
This results in poor voltage regulation, generator over-loading, circulating currents and instability \cite{ref2}.

\begin{figure}[!tb]
\centering
  \includegraphics[width=0.95\columnwidth]{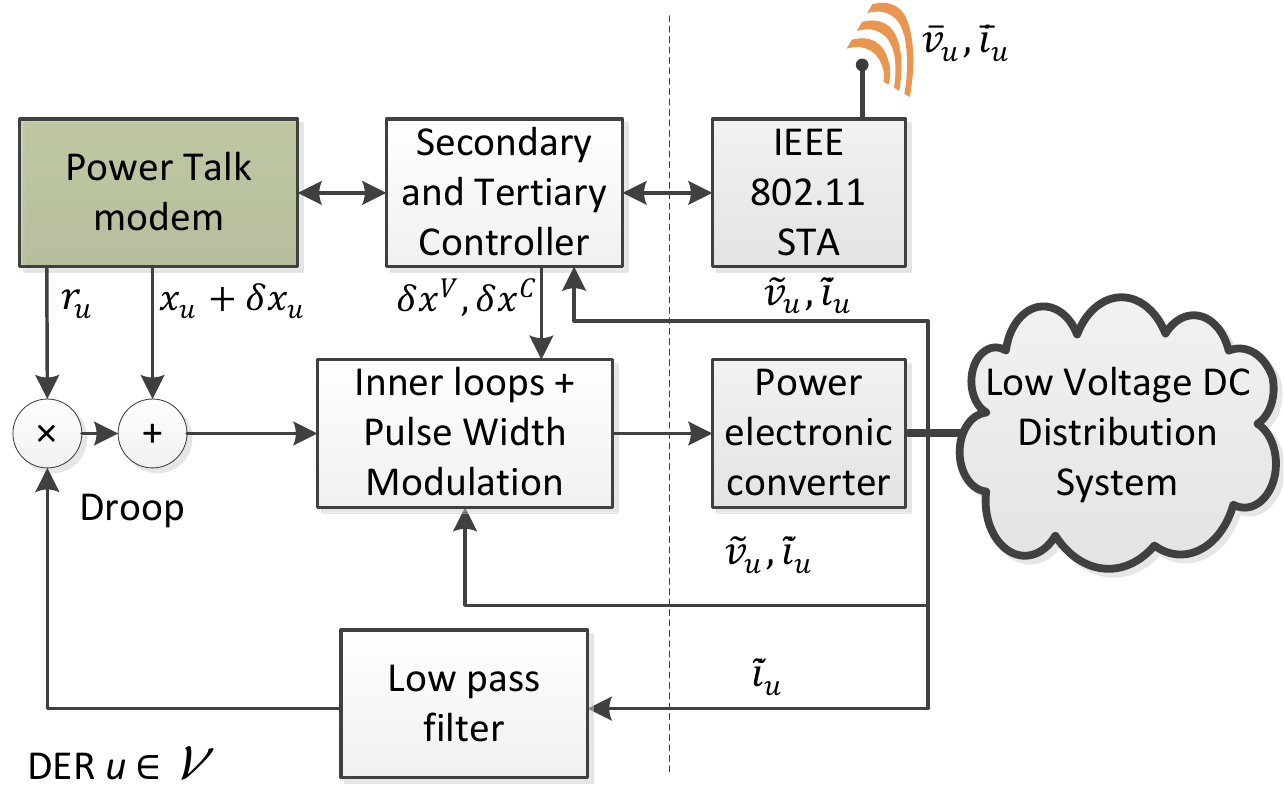}
    \caption{Control diagram and information flow in VSC DER equipped with power talk modem.}
    \label{fig:control}
\end{figure}

In this paper we introduce a novel, secure and robust strategy for distributed secondary control based on dynamic reconfiguration of the communication network, using an auxiliary low-rate powerline communication (PLC) interface that is based on \emph{power talk} \cite{ref5}, \cite{ref7}, \cite{ref6}.
Power talk is a recent communication technique where the information messages are modulated into the parameters of the primary droop control loops of voltage source controllers (VSC), see Fig.~\ref{fig:control}.
This incurs steady state voltage disturbances from which the information can be demodulated by remote controllers.
Unlike existing PLC standards, power talk does not require installation of dedicated hardware, as it is implemented by software modifications of the primary control loops.
In our solution, each controller monitors the wireless channel 
and uses power talk to periodically broadcast a list of peer controllers from which it can receive information over the WiFi interface.
The power talk broadcasts are received by all other controllers in the MG, irrespective whether the controller operates as VSC and actively participates into secondary control, or current source controller (CSC) that does not participate in secondary control.
When a VSC detects that the wireless interface is under DoS attack, it notifies the remaining controllers by sending an empty list, which triggers a graph reconfiguration mechanism.
Specifically, the jammed VSCs switch to CSC mode and leave the secondary control, while one or more CSCs switch to VSC mode and join the secondary control.
The PLECS\textsuperscript{\textregistered{}} based simulation results show that the secondary control regulation can be successfully maintained after DoS attack.

The proposed solution can be seen as a specific instance of more general \emph{software-defined} communication-control framework \cite{ref17}.
In particular, the MG agents use simple software upgrades of the secure and reliable power electronic equipment as a side, low-rate communication channel to exchange control (i.e., power talk) data for network reconfiguration, splitting the network control from the data plane and increasing the overall MG resilience and performance.
To the best of our knowledge, the present paper is the first work that applies the software-defined networking paradigms in the context of MG system with high penetrations of power electronic converters.


The rest of the paper is organized as follows.
Section~\ref{sec:security} reviews the DoS attacks against IEEE 802.11 networks, as well as jamming detection mechanisms.
Section~\ref{model} introduces the system model.
Section~\ref{sec:strategy} presents the proposed reconfiguration strategy against DoS attacks.
Section~\ref{sec:results} verifies the viability of the strategy through PLECS\textsuperscript{\textregistered{}} -based simulations. 
Section~\ref{sec:conc} concludes the paper.

\section{Jamming Attacks to IEEE 802.11 Systems}\label{sec:security}

A jamming attack against an IEEE 802.11 system can lead to severe communication outages, even if the jammer is located at a considerable distance from the MG. A malicious attacker equipped with a jammer can try to corrupt control packets of the protocol, as well as emitting noise, e.g., transmiting packets that do not comply with the protocol to prevent the reception \cite{ref8}. If the jammer is able to produce a continuous noise, the effect is the \emph{back-off freezing} at the transmitter side. In this case it may not even possible to find idle slots for the packet transmission.

Many countermeasures to this DoS attack are considered in literature, ranging from the channel hopping to the direct sequence spread spectrum \cite{ref9}. Uncoordinated adaptations for these techniques are also presented for the broadcast networks \cite{ref11}. Also, the use of directional antennas can be considered to mitigate the interference, although they reduce the spatial coverage, i.e., the number of links among nodes \cite{ref10}.

The jamming detection, which can be implemented locally in each device, is based on various metrics, like the energy reported by the Carrier Senser, the Carrier Sensing Time or the packet error probability \cite{ref9,ref8}.

\section{System Model}
\label{model}

\subsection{Grid Model and Control Architecture}

A DC MG comprises $U$ DERs indexed in $\mathcal{U}=\left\{1,...,U\right\}$; the DERs are connected in parallel, jointly supplying a common load with aggregate power demand denoted with $d$.
Further, the DERs support dual control mode capability \cite{ref12}; i.e., they can operate either in voltage source control or current source control mode.
We assume that $V$ DERs from the subset $\mathcal{V}\subset\mathcal{U}$ are operating as VSCs, while the remaining $U-V$ DERs from the set $\mathcal{U} \setminus \mathcal{V}$ are operated as CSCs.

\subsubsection{Voltage source control mode}
The DERs configured in VSC mode are responsible of the voltage regulation and fostering optimized and fair load sharing. 
The control architecture of VSC DERs is organized in a hierarchy \cite{ref2}, comprising the primary, secondary and tertiary levels.
 
The \emph{primary control} regulates the output electrical parameters and balances the supply-demand to guarantee stability based on local output measurements, via the \emph{droop} law \cite{ref2}:
\begin{equation}\nonumber
v_u = x_u - r_{u}i_u,\;u\in\mathcal{V},
\end{equation}
where $v_u$ and $i_u$ are the output voltage and current, $x_u$ is the reference voltage and $r_{u}$ is the virtual resistance.
This implementation is known as \emph{decentralized droop control}.
The value $v_u$ serves as input reference to the inner primary control loops that operate with frequency $f_{\text{pc}}$, equal to the sampling
frequency of the converters' ADC. 
The droop controller controls $x_u$ and $r_u$, where $x_u$ determines the voltage rating of the system, while the $r_u$ determines the load sharing among different DERs.
Note that under primary control, the output voltage varies with load variations.
Moreover, the load current will not be adequately shared among generators as dictated by the virtual resistances, due to mismatched line admittances.

The \emph{secondary control} is slower than the primary control, operating with sampling period $T^{\text{sc}}$, and has two main objectives: (i) upon load variation, it restores the voltage to the nominal value $v^{\star}$, and (ii) it fosters proportional power sharing among generators.
This is achieved by adding correction offsets of the reference voltage control parameter.
Commonly, it is implemented in distributed manner, where each unit implements two proportional-integral (PI) controllers fed by the output of a consensus algorithm \cite{ref13}. For this purpose, in this work we adopt \emph{robust broadcast gossiping} \cite{ref3}, which performs well under communication outages. 
The secondary control operation can be outlined in the following steps: 
\begin{enumerate}
\item in the secondary control period $k$, DER $u\in\mathcal{V}$ broadcasts a packet to the neighboring control agents over the wireless interface with payload $\mathbf{a}_u(k) = [\bar{v}_{u}(k), \bar{i}_{u}(k)]$, representing local estimates of the average voltage and output current, respectively,
\item during the same secondary control period, DER $u$ receives $R$ packets from neighboring agents, which are supplied as input to the robust broadcast gossip algorithm:
\begin{equation}\nonumber
\mathbf{a}_u(k+1) = \left\{
  \begin{array}{lr}
    \beta_u \mathbf{m}_{u}(k) + ( 1 - \beta_u ) \frac{\sum_{j}\mathbf{a}_{j} (k)}{R}, & R>0\\
    \mathbf{m}_{u}(k), & R=0
  \end{array}
\right.
\end{equation}
where $\mathbf{m}_{u}(k)=[\tilde{v}_u(k),\tilde{i}_u(k)]$ is the local measurement vector in secondary control period $k$, $\mathbf{a}_{j} (k)$ is the status of agent $j$ (assumed to be within the communication range of agent $u$) and $\beta_u\in (0, 1)$ is the consensus weight for agent $u$; in this paper we adopt that $\beta_u=V^{-1},\forall u\in\mathcal{V}; $\footnote{Notice that the robustness of this solution is provided by the fact that in the case when $R=0$, the consensus outputs the local measurements $\mathbf{m}_{u}(k)$, instead of the local state vector $\mathbf{a}_{u}(k)$.}
\item the new status $\mathbf{a}_u(k+1)$ is then forwarded to the local PI controllers, together with the measurement vector $\mathbf{m}_u(k)$;
\item the PI controllers generate the reference voltage correction signals $\delta x^{\text{c}}$ and $\delta x^{\text{v}}$, used to modify the local droop controller to restore the output voltage to the rated value $v^{\star}$:
\begin{equation}\nonumber
	v_u^{\star} = x_{u} + \delta x^{\text{c}} + \delta x^{\text{v}} - r_u i_{u},\;u\in\mathcal{V}.
\end{equation}
\end{enumerate}
After a load/generation change, the offsets $\delta x^{\text{c}}$ and $\delta x^{\text{v}}$ converge to stable values when a global consensus is reached.
The addition of these two correction signals to the droop law provides a global voltage regulation reference and fosters proportional load sharing.

Finally, the \emph{tertiary control} optimizes the performance of the system in terms of minimizing the production costs, enables the coordination with other MGs and operates with significantly larger sampling periods $T^{\text{tc}}$, e.g., $5-30$ minutes.

\subsubsection{Current source control mode}

If in CSC mode, the inner voltage control loop is absent and the DER does not participate in voltage regulation.
Rather, the reference for the inner current control loop is usually generated through Maximum Power Point Tracking (MPPT) algorithm to maximize the individual efficiency \cite{ref1}.
It is modeled as a constant current generator (with internal resistance) with injection current $i_{u}^{\text{csc}}$, $u \in \mathcal{U} \setminus \mathcal{V}$, determined by the MPPT.
From the perspective of efficient resource utilization, it is desired that significant portion of the DERs are operated in CSC mode, for most of the time \cite{ref12}.

\subsubsection{Communication for secondary control}
To support secondary control, only VSC units are required to transmit packets containing local measurements, while CSC units need not to.
For the purpose of making the proposed reconfiguration scheme more resilient to malicious DoS attackers, in our model we impose the requirement that CSC units transmit dummy beacons, regularly and with the same frequency as the rest of the VSC units.
This comes at the price of increased wireless channel load and increased probability for collisions; nevertheless, taking into account the duration of the secondary control periods, the low number of DERs in typical MG setups, and the short length of the exchanged messages, the channel load is expected to remain within tolerable levels.
Finally, even in case of congestion, the robust consensus algorithm has been verified to cope very well with large packet drop rates \cite{ref3}.



\subsection{The Wireless Communication Interface}

In our model, each DER is equipped with an IEEE 802.11 interface.
From this perspective, the DERs are WiFi Stations (STA) that form an ad-hoc, multi-hop network that relies on Distributed Coordination Function (DCF) for medium access \cite{ref15}, see Fig.~\ref{fig:solution}.
The communication range of STAs is denoted with $\rho$.
We assume that each DER has at least one neighbor in its communication range; this implies that the graph $\mathcal{G}$ representing the WiFi network is fully connected.
Also, to guarantee convergence of the secondary control, the subgraph $\mathcal{G}_{\mathcal{V}}$, formed by the secondary control agents should be also fully connected. 
Since the packet arrivals in the controllers' transmission queues are synchronized, a random delay $T^{\text{ad}} \in [0, \tau_d]$ with $\tau_d < T^{\text{sc}}$ is added to decorrelate the start of the channel contentions.

\subsection{Motivating Scenario: Jamming Attack}

\begin{figure}[!tb]
\centering
  \includegraphics[width=0.7\columnwidth]{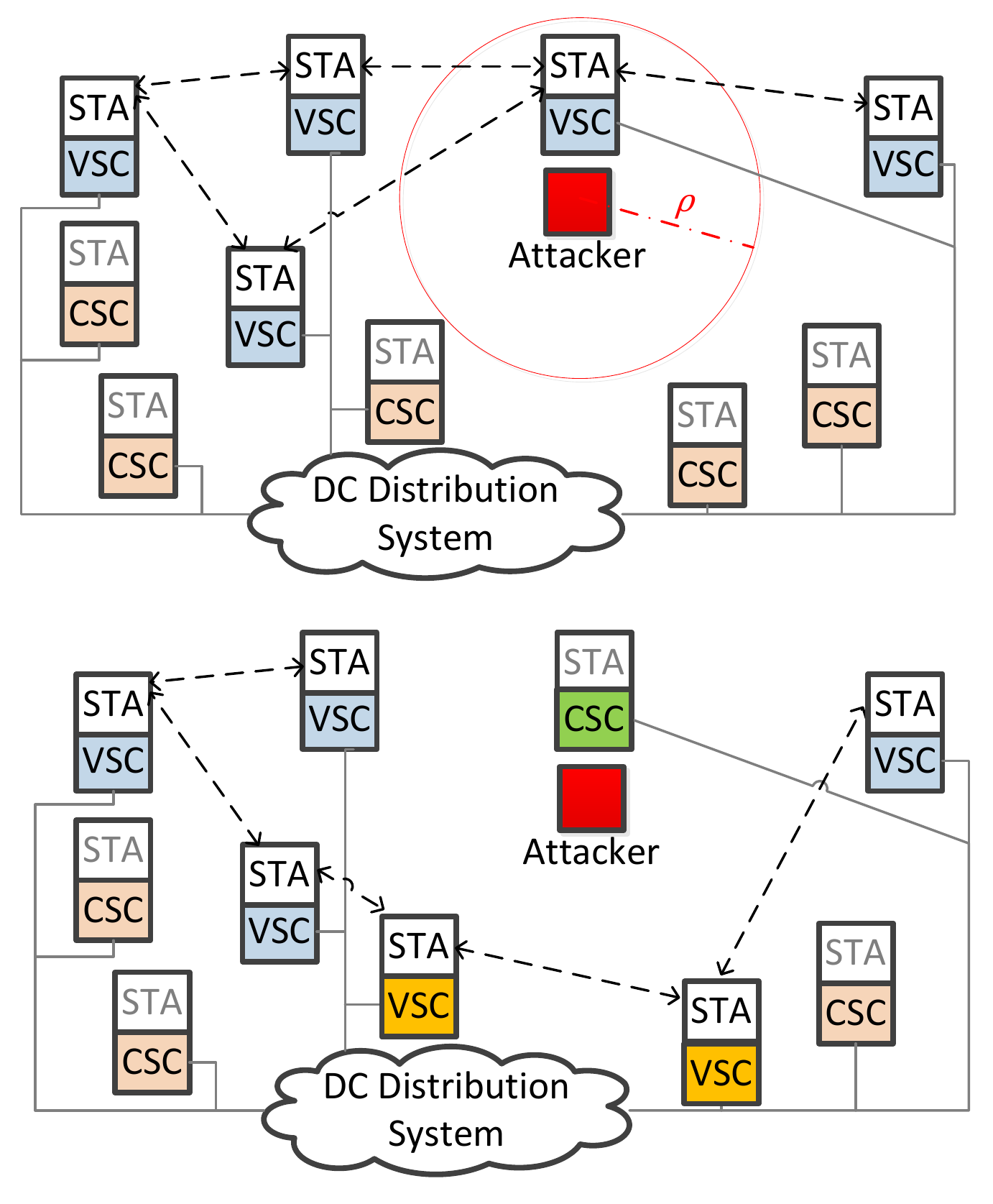}
    \caption{Illustration of the proposed idea (dashed lines represent wireless links, full lines represent powerlines). The blue VSC compose the set $\mathcal{V}$. The jammer is able to split the communication graph. After reconfiguration, the jammed agent is switched to CSC mode (the green CSC), and previously non-participating CSC is switched to VSC mode (the orange VSC), forming the set $\mathcal{V}^{'}$ and reconnecting the graph.}
    \label{fig:solution}
\end{figure}

Assume a malicious attacker using a jamming device capable of emitting a continuous signal on the same frequencies used by the WiFi interface \cite{ref8}.
The jammer has unlimited energy but a limited transmission range equal to $\rho$, see Fig~\ref{fig:solution}.
If a VSC DER is in the attacker transmission range, it can neither transmit packets, because its DCF back-off counter is frozen, nor it is able to receive.
The attack leads to partitioning of the communication graph $\mathcal{G}_{\mathcal{V}}$ into $\omega\geq 1$ subgraphs and loss of global connectivity.
The corresponding subsets of VSCs are denoted with $\mathcal{V}_{\omega}$, such that $\cup_{\omega}\mathcal{V}_{\omega}=\mathcal{V}$.
This disables the secondary control system from performing its tasks properly.
Particularly, the current sharing will be misbalanced, creating generators over-loading and circulating currents.
This situation can be particularly threatening in larger, interconnected systems of MG clusters; loosing the connectivity between the clusters will lead to disproportionate current flows among the cluster, compromising the cost efficiency and stability, and might damage the equipment.


\subsection{Power Talk}
\label{sec:pt}

\begin{figure*}[t]
\centering
{\includegraphics[width=1.9\columnwidth]{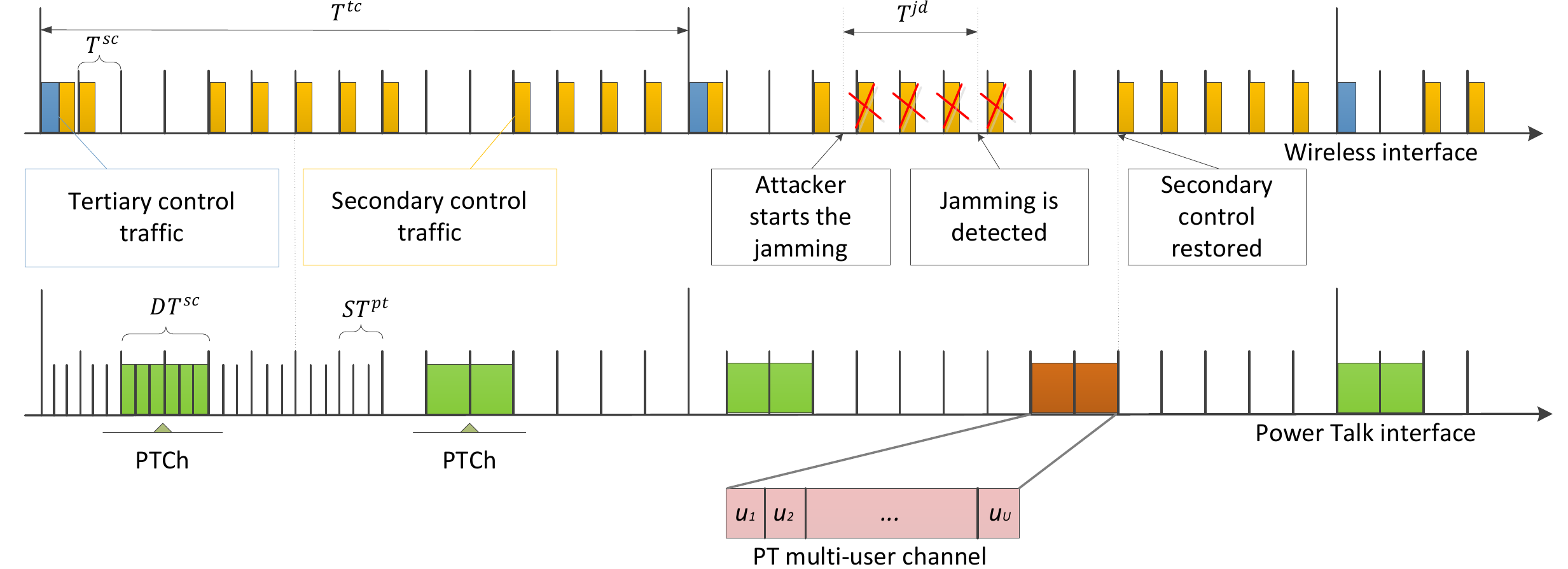}}
    \caption{Time organization of the proposed power talk-based reconfiguration strategy.}
		\label{fig:axis}
\end{figure*}

Power talk is a communication technique that conveys information by varying the droop control parameters of the VSC DERs \cite{ref5}, see Fig. \ref{fig:control}. Therefore, the secondary control of all the VSCs should be disabled during the power talk.
A VSC DER $u$ participating in power talk maps the information symbols in the variations of its local reference voltage $x_u$, resulting in disturbances of the grid voltage that can be measured by all other DERs (both VSCs and CSCs). 
In this paper, we employ binary power talk introduced in \cite{ref5}, where only one DER transmits at a time in TDMA fashion.
Specifically, from power talk perspective, the time axis is slotted into power talk slots of duration $T^{\text{pt}}$ and all communicating DERs are synchronized on slot level.
Label the power talk transmitter in a power talk slot with $u$ and the receiver with $j$.
The information messages, represented as binary strings, are mapped into binary stream of deviations of $x_u$ in the transmitters' droop control loop:
\begin{align}\nonumber
& 0\;\leftrightarrow\;x_{u} - \gamma,\\\nonumber
& 1\;\leftrightarrow\;x_{u} + \gamma,
\end{align}
The deviation amplitude $\gamma$ satisfies $\frac{\gamma}{x_u}\ll 1$, see \cite{ref5} for discussion on how to choose $\gamma$.
Further, these deviations lead to deviations of the output voltage of the receiving DER:
\begin{align}\nonumber
& 0\;\leftrightarrow\;v_{j} - \Delta v_j(0),\\\nonumber
& 1\;\leftrightarrow\;v_{j} + \Delta v_j(1),
\end{align}
where $v_j$ corresponds to $\gamma = 0$.
The receiver collects $f_{\text{pc}} (T^{\text{pt}}-\tau)$ steady state output voltage samples $\tilde{v}_j[n]$ in each slot, compares their average to the threshold $v_j$, and makes decision on the information bits:
\begin{align}\nonumber
\text{if } \frac{1}{f_{\text{pc}} (T^{\text{pt}}-\tau)}\sum_{n}\tilde{v}_j[n]  > v_j \text{, decide}\; 1, \\\nonumber
 \text{if } \frac{1}{f_{\text{pc}} (T^{\text{pt}}-\tau)}\sum_{n}\tilde{v}_j[n]  < v_j \text{, decide}\; 0.
\end{align}
Notice that $\tau$ denotes the transient interval in which the bus reaches a steady state.

The above detection scheme is challenged by two major impairments: (i) the susceptibility of the primary control level to sporadic load variations during the periods in which the secondary control is off, and (ii) sampling noise of the converters' ADC.
Specifically, in the above scheme, a load change invalidates the detection threshold $v_j$, which might lead to burst of bit errors.
A simple strategy to deal with this impairment is to employ load change detection (by tracking the output voltage level), in parallel with power talk symbol transmission/detection, both in the transmitter and the receiver \cite{ref5}. 
After detecting load change, the transmission is paused, $M$ ``blank'' power talk slots, i.e., slots with $\gamma = 0$, are inserted and the communicating DERs measure the new steady state output voltage level, which is used as the new detection threshold.
The viability of this technique is verified in Section~\ref{sec:results}.
Further, the sampling noise of the converter follows Gaussian distribution \cite{ref5b}, i.e., $\tilde{v}_j[n]\sim\mathcal{N}(v_j + \Delta v_j(b),\sigma^2)$,
where typical values for $\sigma$ range from $10^{-2}$ to $10^{-4}$ \cite{ref7}.
A channel code can be employed to protect the information messages messages against noise-related errors.

\section{Secure and Robust Anti-Jamming Strategy based on Power Talk}
\label{sec:strategy}

Here we describe the secure and robust graph reconfiguration scheme when the secondary control system is under DoS attack.
The strategy relies on the following idea: when the wireless interfaces of one or more secondary control agents are attacked and the graph $\mathcal{G}_{\mathcal{V}}$ is disconnected, new connected graph, denoted with $\mathcal{G}_{\mathcal{V}^{'}}$ can be formed by reconfiguring the set of secondary control agents; specifically, previously non-participating CSC DERs in $\mathcal{U} \setminus \mathcal{V}$ can be switched to VSC mode and join the reconfigured set of secondary control agents $\mathcal{V}^{'}$, while the jammed agents are switched to CSC mode and join the subset $\mathcal{U} \setminus \mathcal{V}^{'}$. Fig.~\ref{fig:solution} illustrates the described concepts.

In this paper, we adopt the following graph reconfiguration strategy.
We impose to keep $V$ (i.e., the number of VSCs in the system) constant to guarantee the same voltage restoration capability.\footnote{This allows for keeping the consensus algorithm weight fixed, since we adopted $\beta_u=V^{-1},\forall u\in\mathcal{V}$.} 
Each DER constructs $U\times U$ binary matrix $\mathbf{Q}$ where entry $[\mathbf{Q}]_{u,j}=1$ if DER $u\in\mathcal{U}$ can receive from DER $j\in\mathcal{U}$ over the wireless interface and $0$ otherwise.
Finally, using $\mathbf{Q}$ and a predetermined metric matrix $\mathbf{M}$ of dimensions $U\times U$, the new set $\mathcal{V}^{'}$ is constructed.
The matrix $\mathbf{M}$ can be chosen according to different criteria. For instance, VSC units can be chosen such that the link distances are maximized, in order to mitigate the effect of the jammer.\footnote{In fact, enabling all the VSC unit in the same region will make easier to the attacker to block the secondary control.}
If the DER' positions are unknown, the metric can be based on other parameters, e.g., the Received Signal Strength Indicator (RSSI) of the neighbors' packets.

The matrix $\mathbf{Q}$ is constructed by each DER using information broadcasted and received over the power talk interface.
Specifically, the power talk communication protocol is based on periodic intervals named \emph{power talk channels} (PTCh) of duration $T^{\text{PTCh}} = D T^{\text{sc}}$, during which the secondary control is suspended by each agent, as represented in Fig.~\ref{fig:axis}.
In addition, all CSC units switch to VSC control mode to be able to use the power talk transceiver, described in subsection~\ref{sec:pt}, for the duration of the PTCh.
Each PTCh occurs every $L T^{\text{sc}}$, and in each PTCh, the DERs transmit in TDMA fashion, using predetermined number of power talk slots, see the brown slots in Fig.~\ref{fig:axis}.
Particularly, each DER transmits a list of agents that is able to receive information over the wireless interface, i.e., DER $u$ transmits the elements of the $u$-th row of $\mathbf{Q}$ (except for the element $\mathbf{Q}]_{u,u}$ that is by default $[\mathbf{Q}]_{u,u}=1$).
This is accomplished by allocating $S(V-1)$ power talk slots for each DER, with $S\geq 1$.\footnote{Note that in the case when $S=1$, each DER $u$ directly maps $[\mathbf{Q}]_{u,j}$ in the corresponding power talk slot, using uncoded binary power talk modulation. When $S > 1$, the DERs can also use some form of channel coding.}
The total duration of each PTCh is $US(V-1)$.
The jammed secondary control agents send empty lists.
At the end of a PTCh, all agents have knowledge on $\mathbf{Q}$ and can verify the connectivity of $\mathcal{G}_{\mathcal{V}}$.
Specifically, if the number of subgraphs is $\omega\geq 2$, the reconfiguration is triggered.
We emphasize that power talk is not used to substitute the wireless channel in terms of secondary control, due to its limitation of low bandwidth; it only serves as a secure side channel to support the reconfiguration of the wireless communication system; also, note that we set $L \gg 1$ to avoid frequent interruptions of the secondary control, see further discussion in Sec.~\ref{sec:conc}.
The new set $\mathcal{V}^{'}$ is constructed as follows: select the subset $\mathcal{V_{\omega}}$ with the greatest cardinality and add it to $\mathcal{V}^{'}$.
Then iteratively add other agents to $\mathcal{V}^{'}$ until $V' = V$, by choosing among agents that are in the transmission range of those already in $\mathcal{V}^{'}$.
This information is provided by $\mathbf{Q}$ and ensures that $\mathcal{V}^{'}$ is connected (if it exists).
In case that more than one candidate is present, one of them is chosen according to $\mathbf{M}$.
The fact that $\mathbf{Q}$ and $\mathbf{M}$ are known by all agents ensures that they construct the same $\mathcal{V}^{'}$.

\begin{figure}[!tb]
\centering
\subfloat[The graph before the reconfiguration.]{\includegraphics[width=0.5\columnwidth]{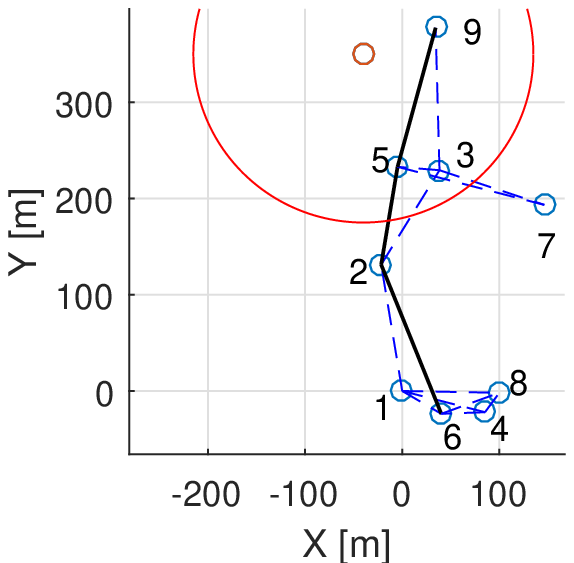}\label{fig:graph_before}}
\hfil
\subfloat[After the reconfiguration.]{\includegraphics[width=0.5\columnwidth]{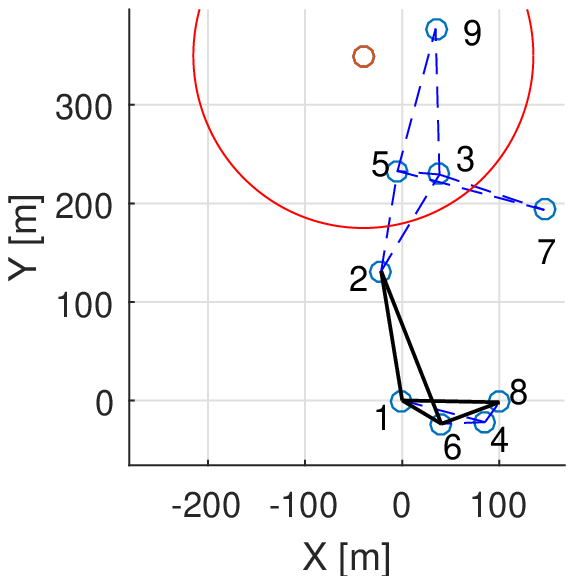}\label{fig:graph_after}}
\caption{Communication graph in the case study. A blue dashed link exists if the distance is lower than $\rho$. A black continuous line links DERs in $\mathcal{V}$.}
\label{fig:graph}
\end{figure}

Finally, DERs decide their operational mode to accomplish the new communication graph.
This operation is also done locally, since DER $u$ can verify if $u \in \mathcal{V}^{'}$ and eventually switch from CSC to VSC mode if $u \notin \mathcal{V}$.
On the contrary, if $u \in \mathcal{V}$, but $u \notin \mathcal{V}^{'}$, $u$ switches from VSC to CSC.
For the other cases, the control mode switch is not necessary.
 

\begin{figure}[!tb]
\minipage{\columnwidth}
\centering
  \includegraphics[width=0.8\columnwidth]{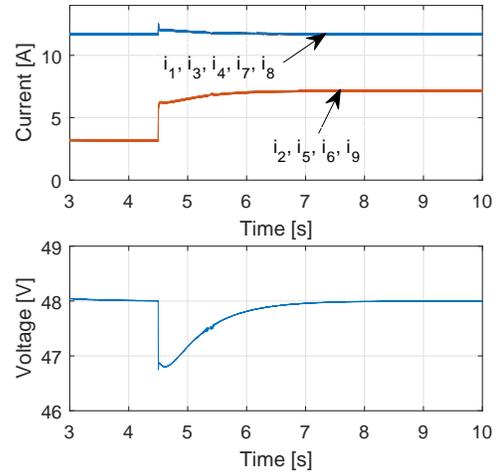}
\endminipage\hfill
    \caption{Simulation of generators' currents and grid voltage when the communication channel is clear and a load variation happens at $t=4.5$ s.}
    \label{fig:no_jam}
\end{figure}

\section{Results}
\label{sec:results}

The verification of the proposed strategy is done by means of a case study that consists of a DC MG operating at the nominal voltage of $v^{\star} = 48 \, \text{V}$. The minimum voltage is $v_{min} = v^{\star} - 10 \% = 43.2 \, \text{V}$, which complies with the standardization proposals \cite{ref16}. The value of all virtual resistances is $r_d = 0.3\,\Omega$ \cite{ref12}, imposing $i_{max} = 13 \, \text{A}$. The $U=9$ DERs are interfaced to the bus by means of buck DC-DC converters, interconnected by a Wi-Fi network and randomly placed on a plane. Specifically, each DG is equipped with a 802.11n wireless interface that operates in ad-hoc mode. 
The $V=4$ DGs operate as VSC, sending UDP packets with a payload of 30 bytes every $T^{\text{sc}} = 0.025 \, \text{s}$. The communication radius  $\rho$ is determined by the path loss, that in our case follows the log-distance model, adopted for the densely populated environments. The path-loss exponent is 3 and the reference loss $46.67 \, \text{dB}$  at $1 \, \text{m}$, resulting in $\rho = 175 \, \text{m}$. The network graph is reported in Fig.~\ref{fig:graph}. 
The described communication network is simulated using ns-3, the controllers using MATLAB/Simulink and the power system modelled in PLECS\textsuperscript{\textregistered{}}.

\subsection{Verification of the Anti-Jamming Strategy}

In the studied case the initial VSC set is $\mathcal{V} = \{ 2, 5, 6, 9\}$, interconnected as shown in Fig.~\ref{fig:graph_before}. DERs belonging to this set participate to the primary, secondary and tertiary control, while the rest of them operate in CSC mode. If a load is plugged to the grid, DERs in $\mathcal{V}$ equally increase their output current and $v$ becomes $v^{\star}$ after the transient time, as reported in Fig.~\ref{fig:no_jam}.\footnote{Note that this result shows that the robust broadcast gossiping algorithm \cite{ref3} can be used for a multi-hop network with random channel access protocol, which is beyond the scenario proposed in \cite{ref3}, where all DERs are within communication distance of each other and round-robin scheduling is used.}

\begin{figure}[!tb]
\minipage{\columnwidth}
\centering
  \includegraphics[width=0.8\columnwidth]{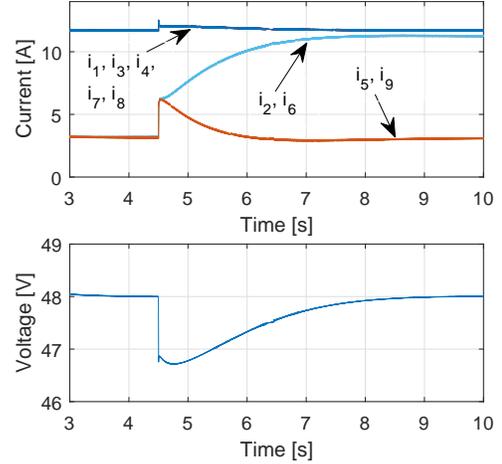}
\endminipage\hfill
    \caption{Simulation of generators' currents and grid voltage when the wireless interfaces are attacked by the jammer.}
    \label{fig:jam}
\end{figure}

In the second scenario, the jammer is positioned in proximity of the grid, see the red node in Fig.~\ref{fig:graph}, and activated at $t = 3 \, \text{s}$. The attack results in the partition of the communication graph in the subsets $\{ 9 \}, \{ 5 \}, \{ 2, 6 \}$. If no countermeasure is taken, the output currents and the grid voltage evolve as in Fig.~\ref{fig:jam}: the two insulated DERs are underutilized, since the last update from neighbors was received under a different load condition, and the currents become unbalanced.

\begin{figure}[!tb]
\centering
{\includegraphics[width=\columnwidth]{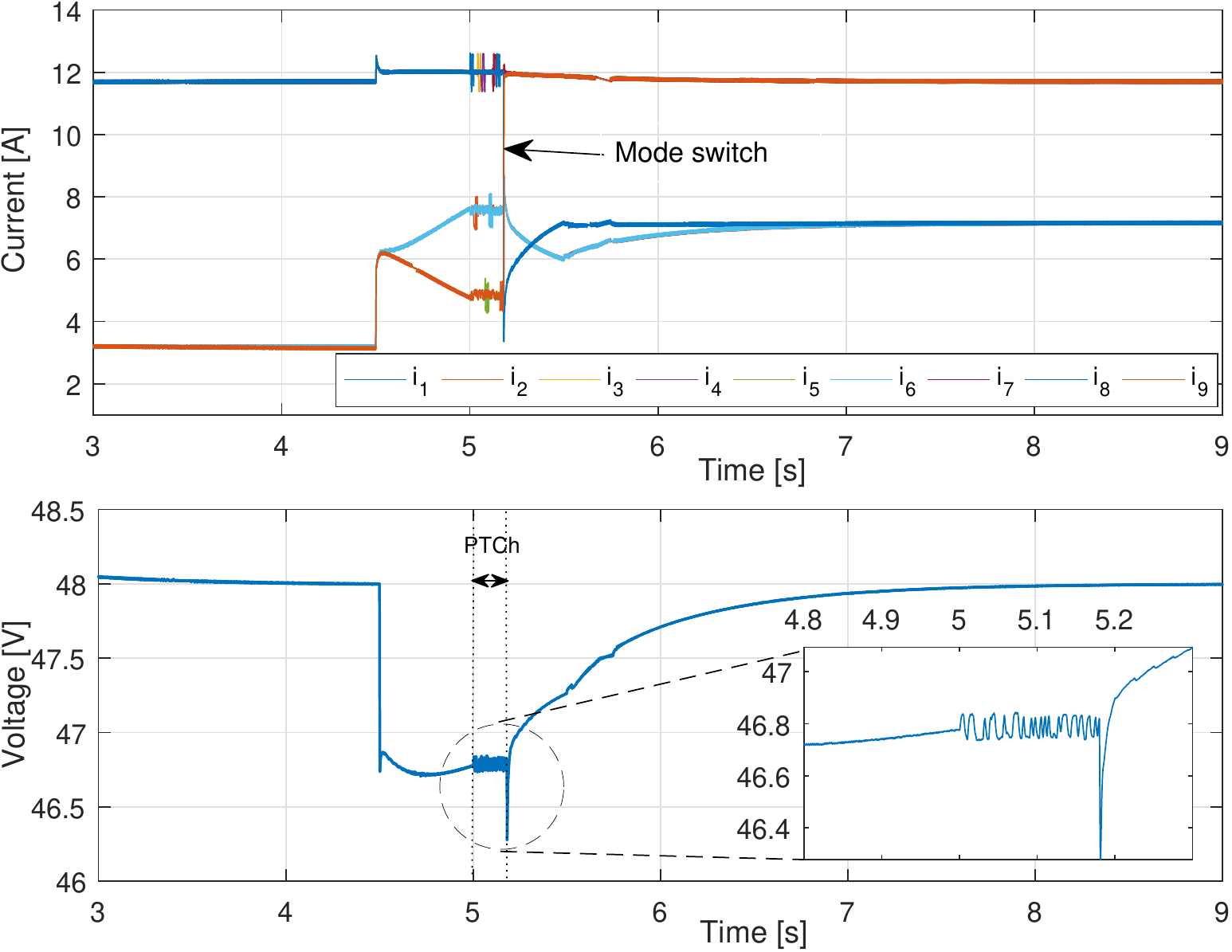}}
    \caption{Simulation of generators' currents and grid voltage when the anti-jamming strategy is executed.}
		\label{fig:plecs_solution}
\end{figure}

Finally, in Fig.~\ref{fig:plecs_solution} we present the result obtained with the adoption of the proposed strategy. The first available PTCh after the jamming detection starts at $t=5 \, \text{s}$ and has duration $T^{\text{PTCh}} = 180 \, \text{ms}$, where the power talk symbol period is $2.5 \, \text{ms}$ and the information transmitted by each agent represented by $V-1$ bits. At the end of PTCh, agents discover that $\mathcal{G_V}$ is partitioned. According to $\mathbf{Q}$, the partition with greatest cardinality is $\{ 2, 6 \}$, that is added to $\mathcal{V'}$, and other connected DERs are $\{ 1, 4, 8 \}$. The selection among these candidates is based on $\mathbf{M}$, which in this example contains the distances between DERs, resulting in a new VSC set $\mathcal{V'} = \{ 1, 2, 6, 8\}$, see Fig.~\ref{fig:graph_after}. The mode switches that happen at $t=5.18 \, \text{s}$ are the following: DERs 1 and 8 from CSCs become VSC, DERs 5 and 9 from VSCs become CSCs. It can be observed that after a transient the currents are effectively equally shared among VSC DERs and the voltage is restored, confirming the validity of the proposed solution.

\section{Discussion and Conclusions}
\label{sec:conc}

In this paper we presented a novel and secure strategy to overcome DoS attacks to the MG distributed secondary control. 
Power talk is used as side communication channel to implement the network control plane, that informs the DERs in the system about the state of the wireless communication network, due to its low complexity and its robustness against malicious attackers that do not have physical access to the grid.
The reliability of the power talk channel makes it attractive and viable side interface for separation of network control and data plane and in the next generation DC MG.
In fact, while the wireless interface provides flexibility at the expense of unreliable or insecure data exchange, the grid itself can propagate the respective meta-data, and increase the overall system resiliency in many different scenarios.

Although the DoS attack was mitigated in the case study, an intelligent jammer may leverage on the knowledge of the physical graph topology to insulate agents located in strategic points.
This may prevent formation of a new $\mathcal{V}'$ with the desired cardinality, resulting in a reduced voltage restoration capability of the system; e.g., for the communication graph of the case study, simultaneous jamming of DERs 1, 2, and 5 prevents formation of any wireless communication graph with $V=4$ DERs.
This kind of scenario can not be mitigated by any algorithm that relies on reconfiguration of the wireless communication graph with a predefined number of involved nodes that is required by the control algorithm.
Nevertheless, in such cases, an alarm message can be transmitted on the power talk channel to a dedicated grid authority.
Furthermore, power talk may be also used in such cases to agree on the reduction of the generated power, i.e., disconnection of part of the CSC DERs, in order to keep the MG in a controllable regime. 
Our ongoing research also aims to evaluate if power talk can entirely replace the wireless interface for disseminating of the secondary control information.
However, because of the low offered bandwidth, new control schemes able to cope with limited and sporadic access to the communication channel are needed.

The proposed strategy can be extended to more sophisticated security threats, such as larger number of (coordinated) jammers that exploit the limitations and vulnerabilities of the physical configuration and topology of the MG as well as the employed wireless technology. In this respect, there is a trade-off between (i) the degrees of the nodes in the graph, (ii) the transmission range and the number of the jammers/STAs, and (iii) the number of agents performing the secondary control $V$.
Its characterization is part of our ongoing research.

Finally, we note that the proposed solution can be generalized to a scenario where the MG is not attacked, but the communication graph is adapted to compensate the variations of the wireless channel. The splitting of network control plane and data plane, together with the active participation of the MG control in the network adaptation, results a promising research topic and can have larger consequences on the MG design.


\section*{Acknowledgment}
The work presented in this paper was supported in part by European Union's Seventh Framework Programme for research, technological development and demonstration under grant agreement no. 607774 ``ADVANTAGE''.

\nocite{*}
\bibliographystyle{IEEEtran}
\bibliography{workshop}

\end{document}